\documentclass[aps,prl,preprint,superscriptaddress,longbibliography]{revtex4-1}
\usepackage{amsmath}    
\usepackage{graphicx}   
\usepackage{bm}

\begin{document}


\title{Deformable ellipsoidal bubbles in Taylor-Couette flow with enhanced Euler-Lagrange tracking}


\author{Vamsi Spandan}
\affiliation{Physics of Fluids and Max Planck-University of Twente Center for Complex Fluid Dynamics, University of Twente, Enschede, PO Box 217, 7500 AE, Netherlands}
\author{Roberto Verzicco}
\affiliation{Dipartimento di Ingegneria Meccanica,University of Rome \lq Tor Vergata\rq, Via del Politecnico 1,
Rome 00133, Italy}
\affiliation{Physics of Fluids and Max Planck-University of Twente Center for Complex Fluid Dynamics, University of Twente, Enschede, PO Box 217, 7500 AE, Netherlands}
\author{Detlef Lohse}
\affiliation{Physics of Fluids and Max Planck-University of Twente Center for Complex Fluid Dynamics, University of Twente, Enschede, PO Box 217, 7500 AE, Netherlands}
\affiliation{Max Planck Institute for Dynamics and Self-Organization, 37077, G\"ottingen, Germany}


\date{\today}

\begin{abstract}
In this work we present numerical simulations of $10^5$ sub-Kolmogorov deformable bubbles dispersed in Taylor-Couette flow (a wall-bounded shear system) with rotating inner cylinder and outer cylinder at rest. We study the effect of deformability of the bubbles on the overall drag induced by the carrier fluid in the two-phase system. We find that an increase in deformability of the bubbles results in enhanced drag reduction due to a more pronounced accumulation of the deformed bubbles near the driving inner wall. This preferential accumulation is induced by an increase in the resistance on the motion of the bubbles in the wall-normal direction. The increased resistance is linked to the strong deformation of the bubbles near the wall which makes them prolate (stretched along one axes) and orient along the stream-wise direction. A larger concentration of the bubbles near the driving wall implies that they are more effective in weakening the plume ejections which results in stronger drag reduction effects. These simulations which are practically impossible with fully resolved techniques are made possible by coupling a sub-grid deformation model with two-way coupled Euler-Lagrangian tracking of sub-Kolmogorov bubbles dispersed in a turbulent flow field which is solved through direct numerical simulations. The bubbles are considered to be ellipsoidal in shape and their deformation is governed by an evolution equation which depends on the local flow conditions and their surface tension.    
\end{abstract}


\maketitle

\section{1. Introduction}
\label{sec:intro}
Deformability of drops or bubbles dispersed in a moving fluid plays a very important role in the overall behaviour of multi-phase flows. The extent of deformability which is determined by the competing actions of viscous, inertia and surface tension forces can affect a drop or a bubble's motion (rectilinear, spiral or zigzag) \citep{clift2005bubbles,ern2012wake,mougin2001path,zenit2008path,ellingsen2001rise}, alter the drop's or bubble's migration or clustering patterns \citep{lu2005effect,lu2008effect,dabiri2013transition} and in some cases also the net momentum or heat transfer in a multi-phase flow \citep{van2013importance,verschoof2016bubble,piedra2015numerical,dabiri2015heat}. It is thus of paramount importance to understand the influence of deformability in multi-phase flows given its wide range of applications (for example bubble columns \citep{deckwer1992bubble}, drag reduction in liquefied gas transport and naval industry \citep{latorre2003micro,ceccio2010friction} etc.). 

Numerous experimental studies have successfully quantified the effect of drops and bubbles on the global momentum transfer in a multi-phase flow \citep{ceccio2010friction}. While such studies reach the extremely high Reynolds numbers relevant for industrial applications, they fall short in detailing the exact physical mechanism behind the observed phenomena. This is due to the extremely challenging measurement techniques required to study the local flow conditions such as velocity profiles of the carrier phase and shape, size, orientation, local volume fraction, acceleration etc. of the dispersed phase. It is possible to measure the aforementioned quantities when there is a single drop or bubble immersed in an optically accessible environment. However, in the case of dispersed multiphase flows with millions of such drops and bubbles simultaneously interacting with the carrier fluid and each other, the measurements either become too invasive or are not able to accurately quantify the local flow conditions. In such situations, numerical simulations which fully resolve both the carrier phase and the dispersed phase would come in handy where one has access to the complete flow field and statistics on the shape, orientation, acceleration and velocity of the dispersed phase at every time instant. These simulations are however limited in scale due to the strongly coupled nature of the problem along with the complex algorithms and numerical schemes required to fully resolve the background turbulent flow along with the dispersed phase simultaneously. For example, state of the art direct numerical simulations (DNS) can only handle several hundreds of deforming drops/bubbles in a weakly turbulent carrier flow \citep{unverdi1992front,tryggvason2001front,prosperetti2001physalis,spandan2016parallel}. Additionally, when the size of the dispersed phase ($d_b$) is of the same scale or smaller than the Kolmogorov scale ($\eta_K$), i.e. $d_b \le \eta_K$, fully resolved simulations become unrealistic due to the extremely fine grids required to resolve each drop/bubble. In such cases, two-way coupled Euler-Lagrangian simulations (point-particle approach) can be employed since then the dispersed phase need not be resolved any more but the acceleration of individual particles, bubbles or drops is computed through empirical force correlations \citep{magnaudet2000motion}. 

Over the last few decades point-particle approaches have been used successfully in the field of multiphase flows given its extremely versatile and computationally inexpensive nature. In particular, studies have focussed on understanding the influence of approximately $10^5$ sub-Kolmogorov particles, drops or bubbles on a turbulent carrier flow \citep{maxey1983equation,elghobashi1994predicting,magnaudet2000motion,mazzitelli2003effect,toschi2009lagrangian,chouippe2014numerical,sugiyama2008microbubbly,spandan2016drag,zhang2001ellipsoidal,yin2003modelling,mortensen2008orientation,marchioli2010orientation,feng2013analysis,njobuenwu2015dynamics,voth2017anisotropic}. However, in contrast to fully resolved approaches, point-particle Euler-Lagrange tracking does not allow for deformation or orientation dynamics which limits its versatility. 

From the above discussion it is clear that there is a need for a technique which can handle large numbers ($O(10^5)$) of sub-Kolmogorov drops or bubbles in highly turbulent flows while allowing for deformation and orientation dynamics to be present. To this effect, in this paper we first build upon the currently existing point-particle approach to account for deformability by coupling traditional Euler-Lagrange tracking with a sub-grid deformation model. We also employ two-way coupling in between the dispersed phase and the carrier phase which allows us to study their influence on each other. With this enhanced two-way coupled point-particle approach we are able to simulate approximately $10^5$ continuously deforming sub-Kolmogorov drops or bubbles dispersed in a turbulent flow field. In particular, we focus on the effect of deformability of low-density bubbles on the net friction or drag in a turbulent Taylor-Couette flow. Here it is important to remember that given the sub-Kolmogorov nature of the dispersed phase, numerical simulations of such large-scale multiphase systems which can also account for deformability can only be possible through the enhanced two-way coupled Euler-Lagrange approach introduced in this paper. 

Taylor-Couette (TC) flow is one of the canonical systems used to study turbulence in wall-bounded sheared flows (see \citet{grossman2016high} for a recent review). In TC flow, the fluid is confined between and driven by two independently rotating co-axial cylinders. The strength of the driving can be quantified using $Re_i=r_i\omega_i(r_o-r_i)/\nu$ and $Re_o=r_o\omega_o(r_o-r_i)/\nu$ which are the inner and outer cylinder Reynolds number, respectively ($r_i$, $r_o$ and $\omega_i$, $\omega_o$ are the radii and angular velocities of the inner and outer cylinder, respectively, while $\nu$ is the kinematic viscosity of the carrier fluid). In this study, we keep the outer cylinder stationary (i.e. $Re_o=0$) and allow only inner cylinder rotation. With increasing driving strength, the flow undergoes several transitions from an initially purely laminar azimuthal flow to an eventually fully turbulent flow. Intermediate regimes include Taylor vortex flow, wavy vortex flow, modulated wavy vortex flow etc.; see \citet{andereck1986flow,ostilla2014exploring,fardin2014hydrogen,grossman2016high} for a detailed overview of these regimes. 

Several numerical and experimental studies have demonstrated that injection of a small amount of dispersed phase such as bubbles can lead to strong drag reduction effects in TC flow \citep{murai2005bubble,murai2008frictional,sugiyama2008microbubbly,van2013importance,spandan2016drag}. Here, drag reduction refers to a decrease in the net torque required to keep the cylinders rotating at a specific angular velocity. The extent of drag reduction, flow modification and the physical mechanisms at play depends on the specific flow regime.

When the Reynolds number in the flow is such that boundary layers and bulk of the flow are laminar and turbulent, respectively, the coherent structures such as Taylor rolls play a dominant role in the angular momentum transport. In this regime, sub-Kolmogorov spherical bubbles ($d_b<\eta_K$) are highly effective in reducing the overall drag on the rotating cylinders \citep{spandan2016drag}. However, the drag reduction effects induced by sub-Kolmogorov spherical bubbles reduce drastically with increasing Reynolds numbers. In the highly turbulent regime (i.e. turbulent boundary layers and bulk), experimental studies have shown that {\it large} bubbles ($d_b>>\eta_K$) which can deform are necessary to recover the drag reduction effects observed in the previous smaller $Re_i$ regime \citep{van2013importance,verschoof2016bubble}. Beyond this, there is a lack of clear understanding on the relevant parameters and conditions crucial for strong drag reduction and also the influence of the shape, size, orientation and accumulation of the dispersed bubbles/drops on the flow.

\begin{figure}
\includegraphics[scale=1.0]{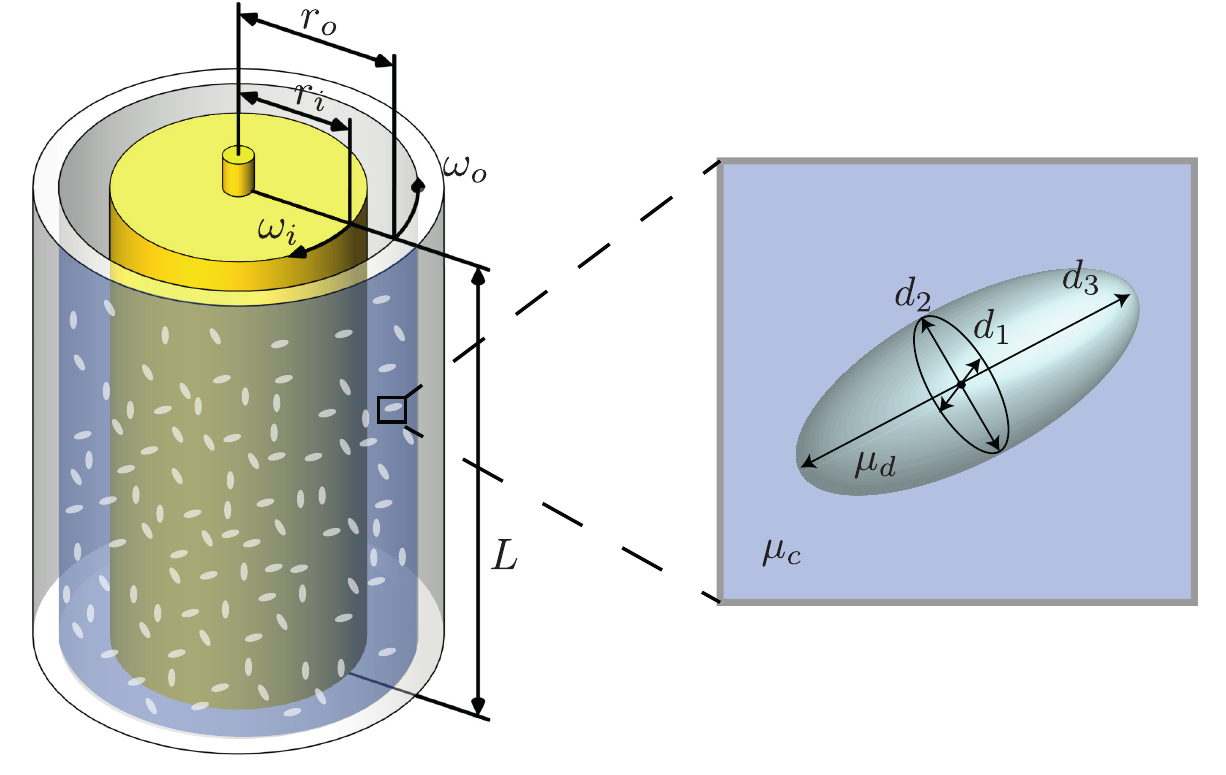}
\caption{Schematic of the Taylor-Couette set-up with a tri-axial ellipsoidal dispersed phase. The shape of the dispersed phase is characterised by the three semi-axes $d_1$, $d_2$ and $d_3$ where $d_1<d_2<d_3$. $\mu_d$ and $\mu_c$ are the dynamic viscosities of the dispersed phase and carrier phase, respectively. For this study the outer cylinder is stationary, $\omega_o=0$}. 
\label{fig:ell_tc}
\end{figure} 

In figure \ref{fig:ell_tc} we show a schematic of the flow set-up used in this study. We consider the shape of the dispersed phase to be deformable tri-axial ellipsoids and compute the deformation through a sub-grid phenomenological model which is described in the next section. As mentioned earlier, our focus here is to understand the effect of deformability of sub-Kolmogorov bubbles on the overall drag on the driving cylinder in TC flow. Additionally, we will look at how deformability affects the mean distribution of the bubbles in the domain. We also answer the question of how these bubbles deform and orient in various regions of the flow (boundary layers and bulk) and how this impacts their effect on the net driving torque in the flow. 

The paper is organised as follows: In the next section we will detail the enhanced two-way coupled Euler-Lagrangian approach which brings together various numerical ingredients to allow for simulations of $10^5$ deformable ellipsoidal bubbles. In section 3 we discuss the results obtained from our simulations and subsequently provide a summary and outlook of the paper in section 4. 

\section{2. Governing Equations}
\label{sec:gov}

\subsection{Dynamics of carrier phase}
The dynamics of the carrier phase is solved using direct numerical simulations (DNS) of the Navier-Stokes equations in cylindrical coordinates. The governing equations read as follows:

\begin{equation}
\frac{\partial \bar u}{\partial t}+\bar u \cdot \nabla \bar u=-\nabla p+\nu\Delta \bar u + \bar f_b(\bar x,t),
\label{eqn:ns}
\end{equation}
\begin{equation}
\nabla \cdot \bar u=0 .
\label{eqn:con}
\end{equation} 
$\bar u$ and $p$ are the carrier phase velocity and pressure, respectively, while $\bar f_b(\bar x,t)$ is the back-reaction force from the dispersed phase onto the carrier phase and its formulation is described later. 
A second-order accurate finite-difference scheme with fractional time stepping is used for the spatial and temporal discretisation of equations (\ref{eqn:ns}) and (\ref{eqn:con}) \citep{verzicco1996finite,van2015pencil}. A non-uniform grid spacing is employed in the radial (wall-normal) direction and periodic boundary conditions are used in the azimuthal (stream-wise) and axial (span-wise) directions. 

\subsection{Dynamics of shape tensor}

We consider the dispersed phase to consist of deformable tri-axial ellipsoidal bubbles ($\rho_b/\rho=0.001$) while the extent of deformation of an individual bubble depends on the local flow conditions. The scheme to solve the deformation dynamics of the ellipsoidal bubbles is detailed in \citet{maffettone1998equation,biferale2014deformation,spandan2016deformation} and is briefly described here for self-consistency. The deformation is computed through the time integration of an evolution equation of a shape tensor $\bar{\bar S}$ \citep{maffettone1998equation}. $\bar{\bar S}$ is a second-order positive-definite symmetric tensor which describes the shape of an individual ellipsoidal bubble, i.e., it satisfies the condition $\bar{\bar S}^{-1}: {\bar x \bar x}=1$, where $\bar x$ is the position vector of any point on the ellipsoid surface relative to its centre. For any known $\bar{\bar S}$, the eigenvalues of the shape tensor give the square of the semi-axes of the ellipsoid while the eigenvectors give the orientation of the semi-axes of the ellipsoid. In non-dimensional form, the sub-grid deformation model reads \citep{maffettone1998equation,biferale2014deformation,spandan2016deformation}:<

\begin{equation}
\frac{d\bar{\bar S}^*}{dt} - Ca(\bar{\bar \Omega}^*\cdot \bar{\bar S}^*-\bar{\bar S}^*\cdot \bar{\bar \Omega}^*)=-f_1(\bar{\bar S}^*-g(\bar{\bar S}^*)\bar{\bar I})+f_2Ca(\bar{\bar E}^*\cdot \bar{\bar S}^*+\bar{\bar S}^* \cdot \bar{\bar E}^*).
\label{eqn:deqn}
\end{equation}
Here $\bar{\bar S}^* = \bar{\bar S}/R^2$ is the non-dimensional form of the shape tensor $\bar{\bar S}$. The normalisation uses the radius $R$ of the undistorted volume equivalent spherical shape.  The strain-rate and vorticity-rate tensors are non-dimensionalised using $G=1/\tau_\eta$ (inverse Kolmogorov turbulent time scale) as $\bar{\bar E}^*=\bar{\bar E}/G$ and $\bar{\bar \Omega}^*=\bar{\bar \Omega}/G$. This gives rise to the control parameter $Ca$ in equation (\ref{eqn:deqn}) which is the capillary number and is the ratio of the viscous forces acting on an individual bubble relative to its inherent surface tension forces. The capillary number can be expressed as the ratio of the interfacial relaxation time scale $\tau=\mu_cR/\sigma$ to the characteristic flow time scale $\tau_\eta$ i.e. $Ca=\tau/\tau_\eta = \mu RG/\sigma$. When $Ca<<1$ the deformation dynamics are decoupled from the turbulent fluctuations in the carrier flow and the influence of the instantaneous deformation on the flow can be neglected. In such a case the forcing from the dispersed phase onto the carrier phase ($\bar f_b(\bar x,t)$ in equation (\ref{eqn:ns})) is computed as $\bar f_b(\bar x,t)=\sum_{i=0}^{N_b} \left(\frac{D\bar u}{Dt}-\bar g\right) V_p\delta (\bar x -\bar x_b(t))$; $N_b$ is the total number of bubbles dispersed into the flow, $V_b$ is the volume of an individual bubble and $\bar x_b(t)$ its position \citep{prosperetti2007computational,mazzitelli2003effect,sugiyama2008microbubbly,spandan2016drag}. The function $g(\bar{\bar S}^*)=3III_s/II_s$ in equation (\ref{eqn:deqn}) ensures volume conservation of each bubble ($II_s$ and $III_s$ are the second and third invariant of the shape tensor, respectively); $f_1$ and $f_2$ are parameters which depend only on the viscosity ratio $\hat \mu = \mu_d/\mu_c$; $f_1=40(\hat \mu+1)/((2\hat \mu+3)(19\hat \mu+16))$ and $f_2=5/(2\hat \mu+3)$. For additional details on the function $g(\bar{\bar S}^*)$ and the formulations of $f_1$, $f_2$ the reader is referred to \citet{maffettone1998equation,biferale2014deformation,spandan2016deformation}. 

\subsection{Modelling of effective forces}

We now describe the governing equations used to compute the hydrodynamic forces acting on the ellipsoidal bubbles. In Euler-Lagrangian tracking the forces acting on the dispersed phase depend on the local flow conditions in the carrier phase and are computed through empirical correlations. Here it is important to point out that the formulation of the forces acting on a deformable tri-axial ellipsoidal particle in a generalised flow field is highly non-trivial. The forces depend on a large number of parameters such as the instantaneous particle Reynolds number, lengths of the three semi-axes, relative orientation of the semi-axes with the flow direction, time scale of deformation, anisotropic added-mass effects etc. to name a few. When the ellipsoids are rigid, a number of studies starting from the seminal work of Jeffery \citep{jeffery1922motion} have tried to predict the translational and rotation dynamics of the ellipsoids under given flow conditions \citep{brenner1963stokes,ganser1993rational,ford1998forces,zhang2001ellipsoidal,blaser2002forces,yin2003modelling,loth2008drag,mortensen2008orientation,feng2013analysis,feng2013non,njobuenwu2015dynamics,fan2000wall,ouchene2015drag}. The modelling is further simplified when the ellipsoids are symmetric (i.e. prolate or oblate) and two of the semi-axes are approximately equal in length. 

In this work we adopt the formulations used by Yin \emph{et al.} \citep{yin2003modelling} and more recently by Njobuenwu \emph{et al.} \citep{njobuenwu2015dynamics} who studied the motion of anisotropic particles in unsteady non-uniform flow conditions. While the correlations for drag and lift used in the above mentioned studies are applicable for rigid ellipsoidal particles, they can be adopted for deforming ellipsoidal bubbles under certain conditions. The first condition is that the bubble interface is fully contaminated with impurities or surfactants so that the interfacial boundary condition is no-slip which is similar to that of the rigid particles 
\footnote{Extension to slip boundary conditions is possible, so it is not a strict condition.}.
The second condition is that the bubbles are sub-Kolmogorov and the capillary number $Ca<<1$, i.e. the characteristic flow time scale is much larger than the interfacial relaxation time scale. In such a case the instantaneous drag and lift force experienced by the ellipsoidal bubble would be approximately similar to that of the rigid particle. Additionally, the correlations used by Yin \emph{et al.} \citep{yin2003modelling} and Njobuenwu \emph{et al.} \citep{njobuenwu2015dynamics} are primarily for axis symmetric ellipsoids, while the shape tensor solved in equation (\ref{eqn:deqn}) corresponds to a tri-axial ellipsoid. In order to circumvent this problem we keep the maximum capillary number of the dispersed phase limited to $Ca=0.1$. By limiting the capillary number we ensure that the deformed ellipsoids are close to being axisymmetric  and this will be shown in more detail in the next section. Under these conditions, the momentum equation for the dispersed phase which consists of contributions from drag, lift, added mass and buoyancy can be written as follows: 

\begin{eqnarray}
\rho_p V_p\frac{d\bar v}{dt} = 0.5\rho A_DC_D|\bar u-\bar v|(\bar u-\bar v) &+& 0.5\rho A_L C_L \frac{\hat{e}_3 (\bar u-\bar v)}{|\bar u-\bar v|}(\hat{e}_3 \times (\bar u-\bar v))\times (\bar u-\bar v) \nonumber \\ 
&+&  \rho V_p C_A \left(\frac{D\bar u}{Dt}-\frac{d\bar v}{dt} \right) + \rho V_p \left (\frac{D\bar u}{Dt}-\bar g \right)+\rho_p V_p\bar g
\label{eqn:ell-eqn}
\end{eqnarray}
The first expression on the right hand side of equation (\ref{eqn:ell-eqn}) accounts for the drag force, the second for the lift force and the last two represent the added-mass force and buoyancy, respectively. In equation (\ref{eqn:ell-eqn}), $A_D$ and $A_L$ are the projected areas normal to the direction of drag and lift, respectively while $\bar u$ and $\bar v$ are the velocities of the fluid and the bubble, respectively. The projected areas are computed as $A_D=\pi R^2 (\text{cos} ^2 \alpha +(4d_{31}/\pi)^2 \text{sin}^2 \alpha)^{1/2}$ and $A_L=\pi R^2 (\text{sin} ^2 \alpha +(4d_{31}/\pi)^2 \text{cos}^2 \alpha)^{1/2}$; $\alpha$ is the angle between the largest semi-axes and the relative slip velocity ($\bar u-\bar v$) and $d_{31}=d_3/d_1$ is the ratio between the largest ($d_3$) and smallest ($d_1$) semi-axes. 

The most important components of equation (\ref{eqn:ell-eqn}) are $C_D$ and $C_L$ which are the drag and lift coefficients, respectively and are modelled as follows \citep{yin2003modelling,njobuenwu2015dynamics}: 

\begin{equation}
\frac{C_D}{K_2}=\frac{24}{Re_bK_1K_2}(1+0.118(Re_bK_1K_2)^{0.6567})+\frac{0.4305}{1+\frac{3305}{Re_bK_1K_2}}
\label{eqn:cd}
\end{equation}

\begin{equation}
C_L=C_D\text{sin}^2 \alpha \cdot \text{cos}\alpha
\label{eqn:cl}
\end{equation}
$Re_b=2R|\bar u -\bar v|/\nu$ is the bubble Reynolds number calculated based on the initial undistorted radius of the bubble. $K_1=\frac{1}{3}d_n/(2R) + \frac{2}{3}\psi^{-0.5}$ and $K_2=10^{1.8148(-log \psi)^{0.5743}}$ are coefficients which depend on the particle sphericity $\psi$ which is the ratio of the surface area of an undeformed particle to the surface area of the deformed particle. $d_n=(4A_D/\pi)^{1/2}$ is the diameter of a circle having the same projection area as that of the deformed particle. While there are several formulations of the lift force and the lift coefficient, here we choose the one proposed by Hoerner \citep{hoerner1965fluid} which assumes that the lift is proportional to the drag force and depends on the relative orientation of the ellipsoid major-axis with the slip velocity and the lift coefficient is computed as give in equation (\ref{eqn:cl}). 

The added mass coefficients are computed based on the ratio of the semi-axes $d_{31}$ as computed by Lai and Mockros for axisymmetric ellipsoidal bubbles \citep{lai1972stokes},

\begin{equation}
C_A = \frac{d_{31}\text{ln}(d_{31}+\sqrt{d_{31}^2 -1}-\sqrt{d_{31}^2 -1}}{d_{31}^2\sqrt{d_{31}^2 -1}-d_{31}\text{ln}(d_{31}+\sqrt{d_{31}^2 -1})}.   
\label{eqn:ca}
\end{equation}
While in principle the added mass coefficient for a tri-axial ellipsoid is a second-order tensor, given the complex and intractable nature of the problem here we consider the simplified form of a fully isotropic coefficient as shown above. 

\subsection{Simulation parameters}

In equation (\ref{eqn:ell-eqn}), we set the density ratio of the dispersed phase as $\hat \rho = \rho_p/\rho = 0.001$ and the viscosity ratio in equation (\ref{eqn:deqn}) to $\hat \mu = \mu_d/\mu_c=0.01$. Two different inner cylinder Reynolds numbers of $Re_i=2500$ and $Re_i=8000$ are considered and the radius ratio and aspect ratio of the set up is fixed to $\eta=r_i/r_o=0.833$ and $\Gamma=4$, respectively. It has been ensured that increasing the aspect ratio beyond the chosen value of $\Gamma=4$ does not influence the results or statistics discussed in later sections. The capillary number of the dispersed bubbles is varied from $Ca=10^{-6}$ which corresponds to the spherical-limit (i.e. the dispersed bubbles are un-deformed and fully spherical) to $Ca=10^{-1}$. A volume fraction of $\alpha_g=0.1\%$ bubbles is used in all simulations and periodic boundary conditions are employed for the bubbles in the azimuthal and axial directions. To treat the collision between the wall and the ellipsoids an elastic-collision model based on the bounding box approach as described in \cite{spandan2016deformation} is used here. 

\section{3. Results}
\label{sec:res}
We now move on to describe the results obtained from the simulations. 
\subsection{Global transport quantities}

\begin{figure}
\includegraphics[scale=1.0]{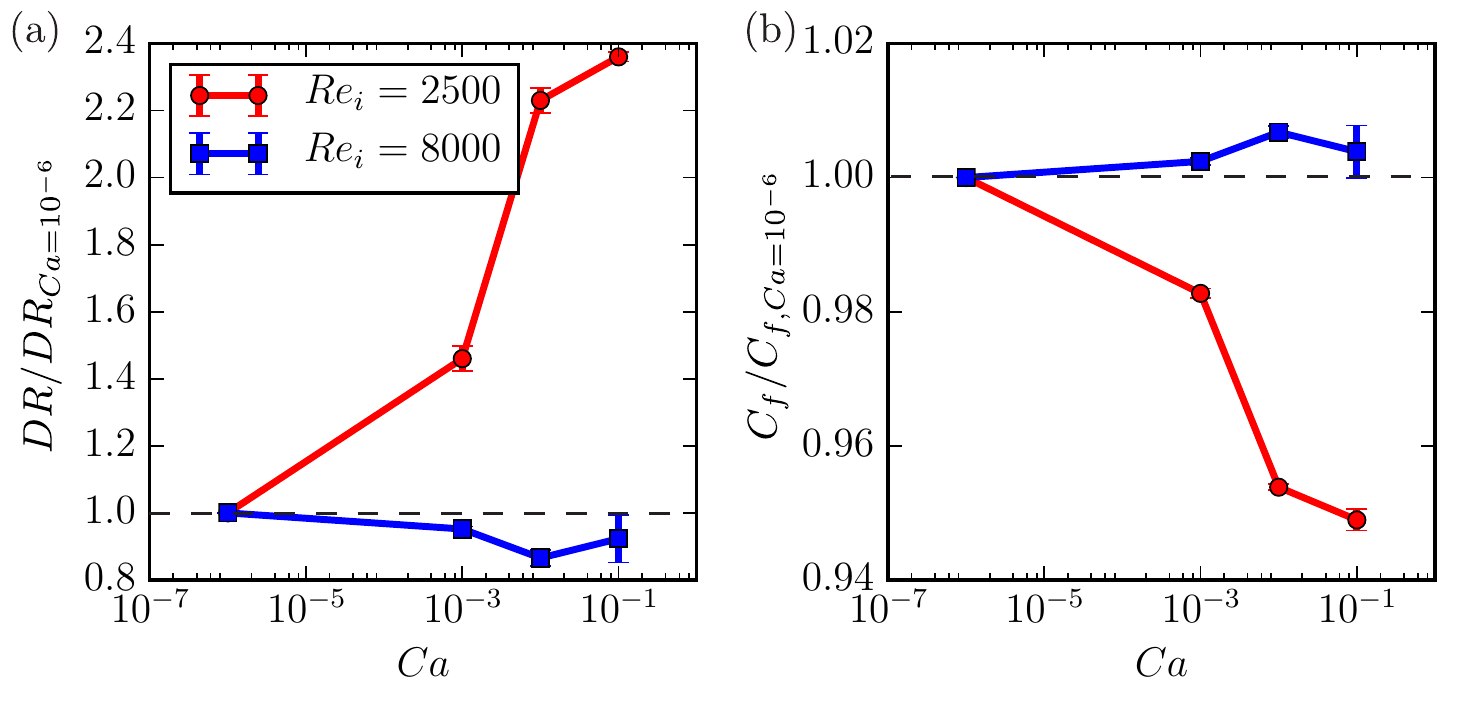}
\caption{(a) Drag reduction ($DR$) and (b) Skin friction coefficient $C_f$ versus Capillary number $Ca$ of the dispersed phase for two different Reynolds numbers}
\label{fig:drcf}
\end{figure}

We will first look at the effect of the bubbles on the global properties of the flow (i.e. the driving torque on the cylinders). In order to do so, we compute the torques required to sustain the flow at a specific Reynolds number in both the single phase ($\tau_s$) and two-phase ($\tau_t$) systems and compute the overall drag reduction (in \%) induced by the dispersed bubbles as $DR=100(\tau_s-\tau_t)/\tau_s$. The driving torques can also be converted into a skin-friction coefficient as $C_f=\sigma_w/0.5\rho U_i^2$, where $\sigma_w$ is the averaged wall-shear stress. Before analysing the effect of deformability of a sub-Kolmogorov dispersed phase in TC flow, it is important to understand the origin of drag reduction in the low capillary number limit, i.e. when the surface tension forces in the bubbles are very large compared to the viscous forces and the bubbles are close to spherical. In such a system it has been shown that the buoyancy of the bubbles disrupt and weaken the plume ejections from the inner cylinder and consequently the Taylor rolls which effectively transfer momentum and this leads to drag reduction \citep{spandan2016drag}. Given that the weakening of the Taylor rolls is just due to the buoyancy of the bubbles dispersed into the flow, we expect that this mechanism is still relevant when the bubbles become deformable. Here, it is important to remember that the system is incompressible and thus deformability has no effect on the overall buoyancy of the bubbles. We now quantitatively compare the drag reduction obtained by spherical and deformable bubbles. 

To understand the effect of deformability, we plot the normalised drag reduction and skin-friction coefficient versus the capillary number of the dispersed phase for two different Reynolds numbers in figure \ref{fig:drcf} (a) and (b), respectively. To show the effect of deformation, both $DR$ and $C_f$ are normalised using their corresponding values at $Ca=10^{-6}$, i.e. in the spherical limit of the bubbles. At $Re_i=2500$, one can observe that with increasing capillary number there is a net increase in drag reduction in comparison to the low Capillary number ($Ca=10^{-6}$) limit. In other words, by weakening the inherent surface tension forces of the sub-Kolmogorov bubbles we are able to more than double the net drag reduction achieved in the flow. In figure \ref{fig:drcf} (b), we plot the same data but in the form of the normalised skin-friction coefficient which decreases with increasing $Ca$ for $Re_i=2500$. At this Reynolds number it is clear that the drag reduction achieved in the low Capillary number limit can be further enhanced by making the dispersed particles (drops or bubbles) deformable. At $Re_i=8000$, there is barely any significant change in the drag with increasing deformation. When sub-Kolmogorov spherical bubbles are used at such high Reynolds numbers, negligible drag reduction has been observed as shown in the studies by \citet{spandan2016drag,murai2005bubble,sugiyama2008microbubbly}. At $Re_i=8000$, the coherent structures lose importance in the angular momentum transport and the turbulent fluctuations start taking over. In such cases, the buoyancy of the dispersed phase is not strong enough in comparison to velocity fluctuations such that it can alter the flow structures and reduce dissipation to achieve drag reduction. We observe in figure \ref{fig:drcf} that then deformability of the dispersed phase has no significant effect on the net drag reduction achieved by a sub-Kolmogorov dispersed phase at high Reynolds numbers. This is justified given the incompressibility of the two-phase system and thus deformability has no effect on the overall buoyancy of the dispersed phase. 

\begin{figure}
\includegraphics[scale=1.0]{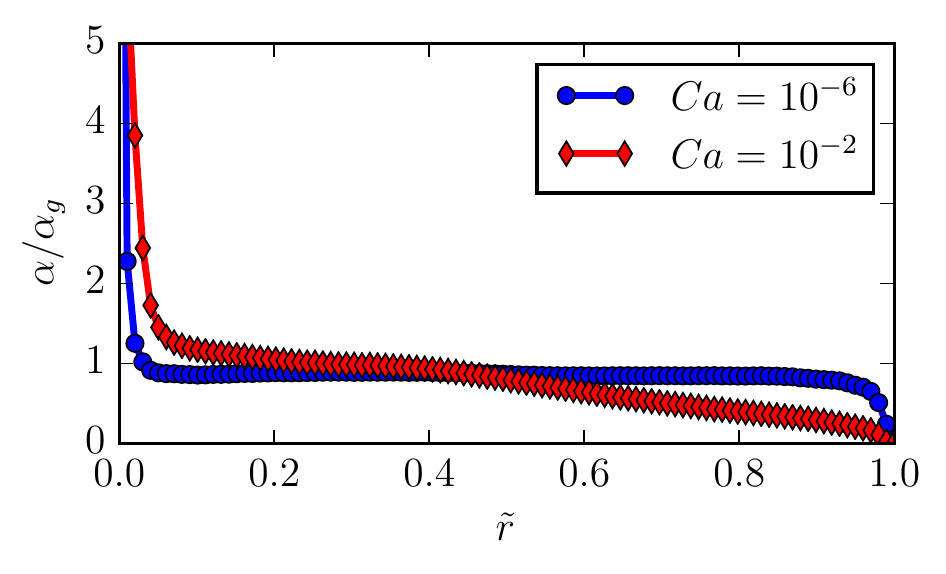}
\caption{Radial (wall-normal) profiles of azimuthally, axially and time averaged local volume fraction of the dispersed phase for two different Capillary numbers at $Re_i=2500$. The profiles for $Ca=10^{-3}$ and $Ca=10^{-1}$ lie below and above the profile of $Ca=10^{-2}$, respectively and are not shown here for clarity.}
\label{fig:rvol}
\end{figure} 

\subsection{Mean bubble concentration profiles}

We now focus on the simulations at $Re_i=2500$ to understand the increase in drag reduction with deformation. In figure \ref{fig:rvol} we plot the radial (wall-normal) profiles of the normalised mean local volume fraction of the dispersed phase. The wall-normal position is normalised using the gap-width as $\tilde r = (r-r_i)/d$ ; thus $\tilde r=0$ refers to the bubbles close to the inner cylinder while $\tilde r=1$ refers to the bubbles close to the outer cylinder. As has been shown in previous studies \citep{murai2005bubble,sugiyama2008microbubbly,spandan2016drag}, rigid spherical bubbles in TC flow ($Ca=10^{-6}$) tend to accumulate near the inner cylinder as a result of the heavier carrier fluid being pushed away by the centrifugal forces. However, as can be seen from figure \ref{fig:rvol}, the capillary number of the dispersed phase seems to play a very important role in the relative distribution of the bubbles in the domain for $Re_i=2500$. While the distribution is close to homogeneous in the bulk for low capillary number, there appears to be a gradient in the bubble distribution at the higher $Ca$. More bubbles tend to accumulate near the inner cylinder with increasing Capillary number. The enhanced drag reduction with increasing capillary number observed in figure \ref{fig:drcf} can be related to an increase in accumulation of the dispersed bubbles near the inner cylinder. As discussed earlier, when rigid spherical bubbles are injected into the TC system, buoyancy is crucial in disrupting and weakening the plumes ejections which are responsible for the angular momentum transport in the wall-normal direction. A larger concentration of bubbles near the inner cylinder implies a stronger effect of the dispersed phase on the plumes being ejected from the inner cylinder which results in an increase in drag reduction.       

\begin{figure}
\includegraphics[scale=1.0]{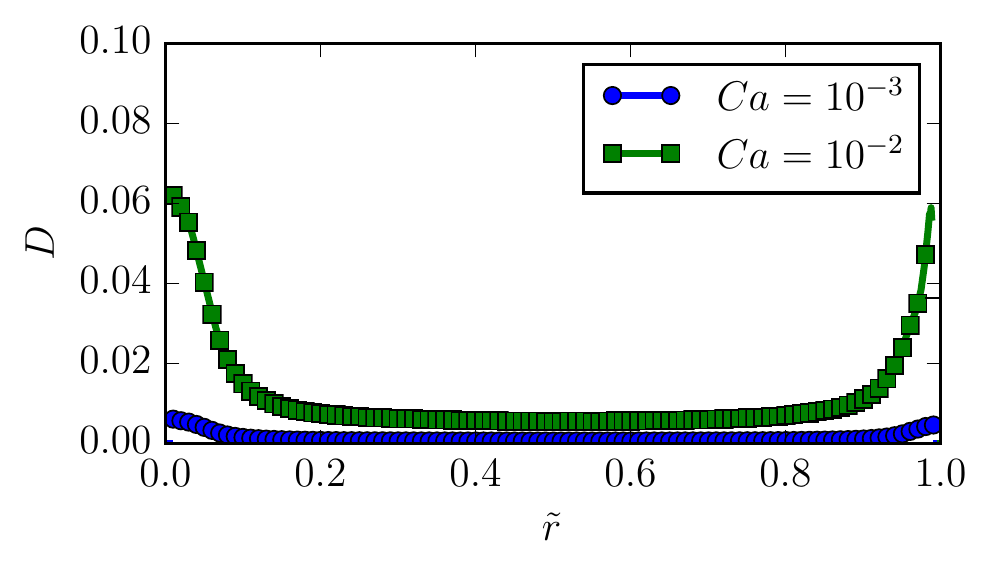}
\caption{Radial (wall-normal) profiles of azimuthally, axially and time averaged deformation parameter of the bubbles for two different Capillary numbers at $Re_i=2500$.}
\label{fig:25rdef}
\end{figure}             

The mean position of the bubbles dispersed in the TC domain is governed by the hydrodynamics forces that act on them which in turn depend on the shape of the bubbles as discussed in the previous section. It thus becomes important to look at the extent of deformation of bubbles in different regions of the flow. For this we use the deformation parameter $D=(d_3-d_1)/(d_3+d_1)$ which gives an idea of the extent of deformation of individual bubbles. Here $d_3$ and $d_1$ are the lengths of the major and minor axis of the ellipsoid, respectively \citep{taylor1932viscosity,taylor1934formation}; for $Ca=0$ we have $D=0$ by definition. In figure \ref{fig:25rdef}, we plot the azimuthally, axially and time averaged profiles of the deformation parameter $D$ versus the normalised wall-normal position $\tilde r$ which shows a clear spatial inhomogeneity in the extent of deformation of the dispersed phase. When the Capillary number is high, the bubbles near the walls deform much more than those in the bulk due to the strong shear in the boundary layers. For both capillary numbers the deformation is relatively lower and homogeneous in the bulk region of the flow which is a result of relatively lower shear provided by the Taylor rolls. Here it is also important to note that the deformation statistics of the bubbles is similar to neutrally buoyant drops dispersed in TC flow \citep{spandan2016deformation}. This is because the deformation is fully governed by equation (\ref{eqn:deqn}) which depends only on the local flow conditions and the inherent surface tension of the dispersed phase and not on buoyancy effects. This is justified given the sub-Kolmogorov nature of the bubbles and that buoyancy effects on the deformation only become relevant when the dispersed phase is much larger than the Kolmogorov scale. 

\begin{figure}
\includegraphics[scale=1.0]{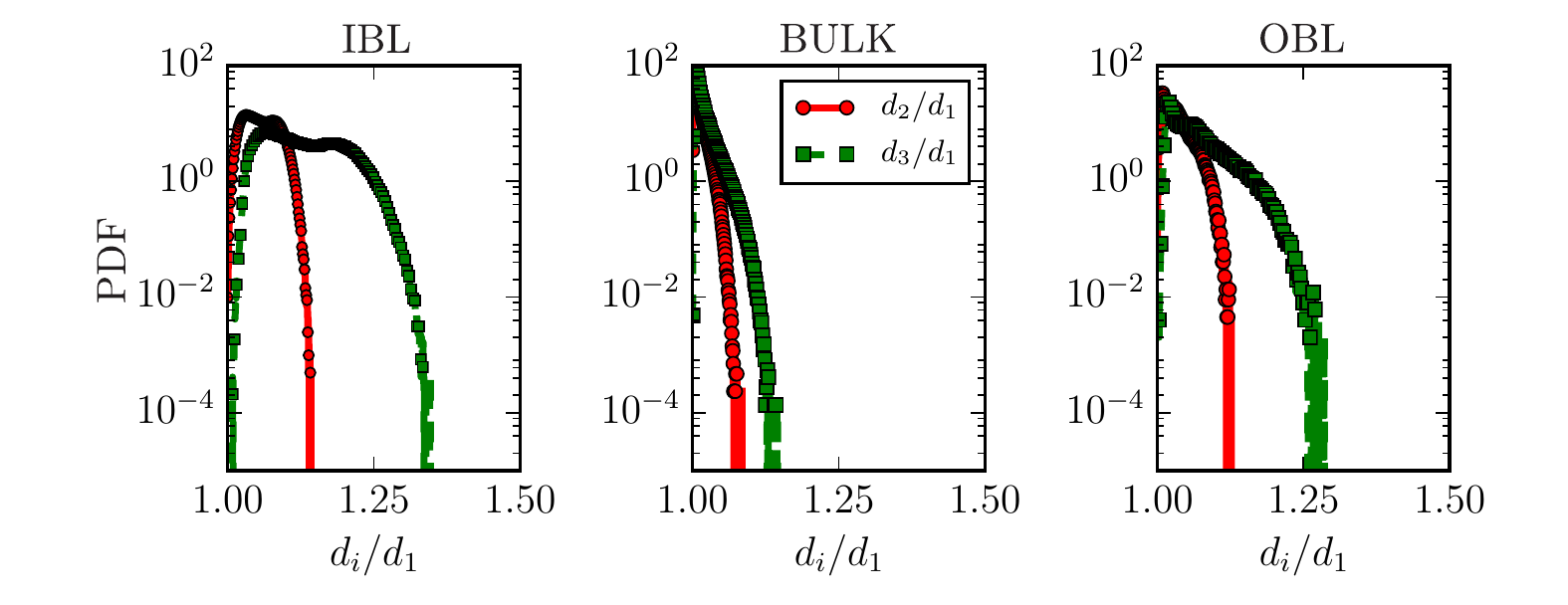}
\caption{Probability distribution functions of the ratio of semi-axes of bubbles in the inner boundary layer (IBL), bulk, and outer boundary layer (OBL) for $Ca=10^{-2}$ at $Re_i=2500$.}
\label{fig:25sem}
\end{figure} 

\subsection {Bubble shape and orientation}

While the deformation parameter gives a good idea on the spatial inhomogeneity of the deformation, it does not give any information on the exact shape of the bubbles in various regions of the flow. In order to understand this more clearly we compute the probability distribution functions (PDF) of the ratios of the ellipsoid axes i.e. $d_2/d_1$ and $d_3/d_1$ ($d_1<d_2<d_3$) in three different regions of the flow, namely the inner boundary layer (IBL), in the bulk and in the outer boundary layer (OBL). These PDF's give us a better idea on whether the dispersed bubbles are more prolate ($d_3/d_1 > 1$ and $d_2/d_1 \sim 1$) or oblate ($d_3/d_1 \sim d_2/d_1 >1$). It has already been seen in the case of neutrally buoyant drops that an increased influence of rotation over stretching can result in oblate shaped drops while prominence of stretching results in more prolate shaped drops \citep{spandan2016deformation}. The PDF's for the case of $Re_i=2500$ and $Ca=10^{-2}$ are shown in figure \ref{fig:25sem}. As can be seen from the distributions in the IBL and OBL, the bubbles are close to being prolate which indicates a stronger influence of stretching over rotation of the bubbles. We also observe from the distribution of $d_2/d_1$ that a large number of bubbles distributed throughout the domain are close to being axisymmetric i.e. $d_2/d_1 \sim 1$. For consistency of our model this is an important observation since the momentum equation used for the dispersed bubbles (equation \ref{eqn:ell-eqn}) is applicable primarily for axi-symmetric particles. When the capillary number of the bubbles goes beyond $Ca=0.1$ the shapes of the dispersed bubbles become fully tri-axial (i.e. all three semi-axes are unequal in length) and in such a case equation (\ref{eqn:ell-eqn}) would need further refinement.     

\begin{figure}
\includegraphics[scale=1.0]{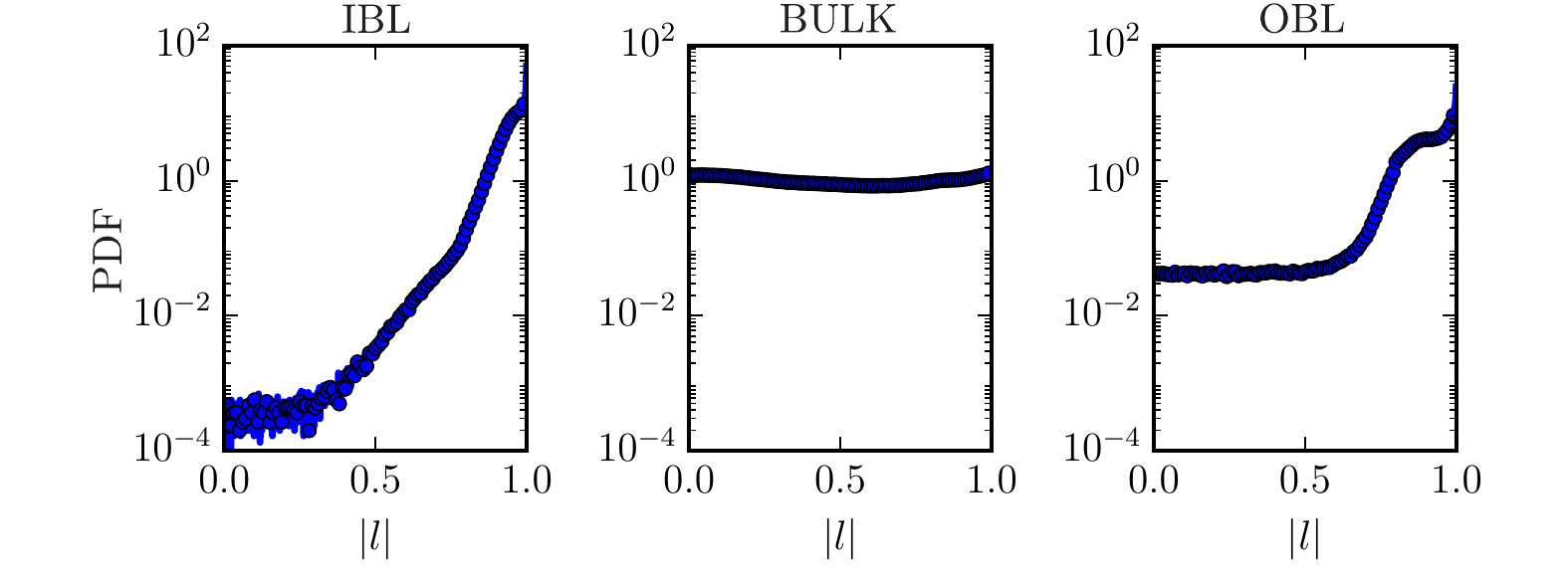}
\caption{Probability distribution functions of the direction cosine of the orientation of the semi-major axes of the bubbles with the azimuthal (stream-wise) direction in the inner boundary layer (IBL), bulk, and outer boundary layer (OBL) for $Ca=10^{-2}$ at $Re_i=2500$.}
\label{fig:25ang}
\end{figure} 

We now look at the distributions of the orientations of the bubbles in the flow. In figure \ref{fig:25ang} we plot the p.d.f of the direction cosine of the orientation of the bubbles with $\hat \theta$ (azimuthal) direction, i.e. $l=\text{cos}(\theta_{\hat e_3 \hat \theta})$ where $\theta_{\hat e_3 \hat \theta}$ is the angle between $\hat \theta$ and the semi-major axis ($\hat e_3$) of an individual bubble. In the IBL we notice that the majority of the bubbles tend to orient in stream-wise direction. A relatively weaker but observable preferential orientation is also found in the OBL while it is highly isotropic in the bulk due to the Taylor rolls. Similar distributions are observed for neutrally buoyant drops dispersed into TC flow \citep{spandan2016deformation} and in the case of solid ellipsoidal particles in a turbulent channel flow \citep{mortensen2008orientation,marchioli2010orientation,njobuenwu2015dynamics}. The orientation of the bubbles with the stream-wise direction is assisted by the stretching provided by the strong shear in the IBL which results in prolate shaped bubbles (c.f. figure \ref{fig:25sem}). Such an alignment preference also minimises the projected area ($A_D$) of the bubble in the direction of the stream-wise drag force. Given the sub-Kolmogorov nature of the dispersed phase the bubble Reynolds numbers $Re_b$ is $O(1)$ in these systems (c.f. figure 12 of \citet{spandan2016drag}). This suppresses any instabilities that may arise in the bubble motion which tends to change its preferential orientation.  

\begin{figure}
\includegraphics[scale=1.0]{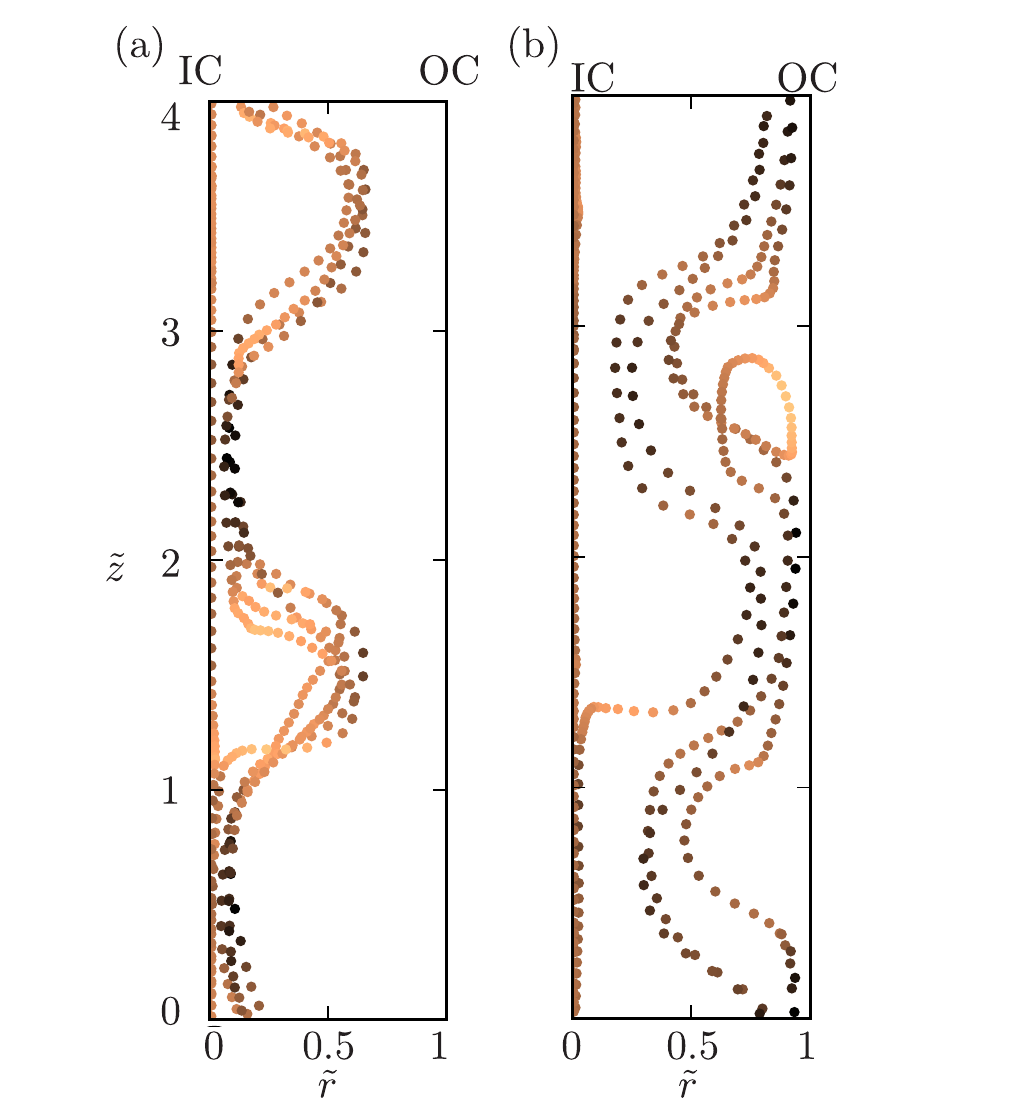}
\caption{Trajectories of randomly chosen bubbles in the $\hat r-\hat z$ plane taken over approximately 100 large eddy turn over times (a) $Ca=10^{-2}$ (b) $Ca=10^{-3}$ at $Re_i=2500$. IC refers to inner cylinder while OC refers to the outer cylinder. The colour scale indicates the axial velocity of the bubble where a dark colour indicates a higher velocity while a lighter colour indicates a lower velocity.}
\label{fig:btraj}
\end{figure}

We now go back to explaining the preferential accumulation of bubbles near the inner cylinder with increasing capillary number as shown in figure \ref{fig:rvol}. From figures \ref{fig:25sem} and \ref{fig:25ang} we observed that the bubbles close to the inner cylinder for $Ca=10^{-2}$ are prolate and also align themselves with the stream-wise direction. While this reduces the overall projected area of the deformed bubbles in the azimuthal (stream-wise) direction it increases the projected area in the radial (wall-normal) direction. To understand the effect of this increased projected area on the overall motion of the bubbles we plot the trajectory of randomly chosen bubbles in the $\hat r-\hat z$ plane for two different capillary numbers in figure \ref{fig:btraj}. We also refer to figure 11 of \citet{spandan2016drag} which shows the effect of buoyancy and Reynolds numbers on the trajectories of buoyant spherical bubbles in TC flow. In both figures, one can clearly see that bubbles enter the bulk from the plume ejection regions near the inner cylinder ($\tilde r=0$ and $\tilde z \sim 1$, $\tilde z \sim 3$ in figure \ref{fig:btraj}) and continue on their upward motion which is assisted by buoyancy. The bubbles moving in the wall-normal direction from the inner cylinder into the bulk are hampered by the increased drag force from the increased projected area which results in their preferential accumulation close to the inner cylinder when they become deformable. As discussed earlier, given the incompressibility of the overall system deformation of the bubbles has no effect on its buoyancy and a larger number of bubbles near the inner cylinder means that the weakening of the plumes by the bubbles is stronger; this in turn leads to stronger drag reduction. 

\section{Summary and Outlook}
\label{sec:summ}

In this work we have brought together direct numerical simulations of Taylor-Couette (TC) flow (a wall-bounded sheared system), force-based Lagrangian tracking of an ellipsoidal dispersed phase and a sub-grid deformation model with which we have improved the conventional Euler-Lagrangian tracking widely used in the study of multi-phase flows. This two-way coupled enhanced Euler-Lagrange approach allows us to track $O(10^5)$ sub-Kolmogorov deformable drops and bubbles in a turbulent flow field and also study their influence on the global and local properties of the flow. The simulations performed using this approach are presently impossible with fully-resolved techniques due to the sub-Kolmogorov nature of the dispersed phase. 

We have used the numerical simulations to understand the effect of deformability of approximately $10^5$ bubbles dispersed into TC flow and in particular we focus on their effect on the overall drag on the rotating cylinder. We find that at an inner cylinder Reynolds number of $Re_i=2500$, increasing the deformability (i.e. the capillary number) results in an increase in the overall drag reduction. On studying the mean bubble volume fraction profiles in the domain, we find that deformable bubbles prefer to accumulate near the rotating inner cylinder. The PDF's of the ellipsoid axes and their orientational preference indicate that near the inner cylinder the bubbles are primarily prolate and tend to align with the stream-wise direction. This results in an increase in the drag experienced by the near-wall deformed bubbles in the wall-normal direction which ultimately results in their accumulation near the inner cylinder. Drag on the inner cylinder is directly connected to the strength of plume ejections which are disrupted more effectively by deformable bubbles due to their preferential accumulation. 

At a relatively higher Reynolds number of $Re_i=8000$ we find that deformability of sub-Kolmogorov bubbles has no effect on drag reduction as the turbulent fluctuations are much stronger in comparison to the buoyancy of the bubbles. At this Reynolds number and beyond, i.e. when both boundary layers and bulk become turbulent, experiments have shown that finite-size (larger than Kolmogorov scale) bubbles which can deform produce strong drag reduction effects in spite of no recognisable difference in their accumulation near the driving wall \citep{van2013importance}. This indicates that the drag reduction induced by finite-size bubbles at very high Reynolds numbers ($Re \sim 10^5-10^6$) is fundamentally different from what has been described in the current work which is more close to the works of \citep{murai2005bubble,murai2008frictional} where $Re_i \sim 10^3$. Additionally, in the case of finite-size bubbles, inertial forces become relevant in the deformation dynamics and this makes the Weber number an important parameter. Numerical simulations of such systems would invariably require fully resolved simulations with methods such as front-tracking, volume of fluid or immersed-boundaries etc. This work is currently in progress \citep{spandan2016parallel}.            

This work was supported by the Netherlands Center for Multiscale Catalytic Energy Conversion (MCEC), an NWO Gravitation programme funded by the Ministry of Education, Culture and Science of the government of the Netherlands and the FOM-CSER program. We acknowledge PRACE for awarding us access to Marconi supercomputer based in Italy at CINECA and NWO for granting us computational time on Cartesius cluster from the Dutch Supercomputing Consortium SURFsara.

\bibliography{../../mylit}

\end{document}